\documentclass[aps,prd,twocolumn,superscriptaddress,amsmath,amssymb,showpacs]{revtex4}

\usepackage{graphicx}
\usepackage{amsmath}
\usepackage{amssymb}
\usepackage{amsfonts}
\usepackage{latexsym}

\usepackage{dcolumn}
\usepackage{bm}

\newcommand\myeq{\mathrel{\overset{\makebox[0pt]{\mbox{\normalfont\tiny\sffamily def}}}{=}}}

\begin{document}

\title{Strong gravitational lensing and black hole quasinormal modes: Towards a semiclassical unified description}

\author{Bernard Raffaelli}
\email{bernard.raffaelli@gmail.com}


\begin{abstract}
We examine in a semiclassical framework the deflection function of strong gravitational lensing, for static and spherically symmetric black holes, endowed with a photon sphere. From a first-order WKB analysis near the maximum of the Regge-Wheeler potential, we extract the real phase shifts from the S-matrix elements and then we derive the associated semiclassical deflection function, characterized by a logarithmic divergent behavior. More precisely, using the complex angular momentum techniques, we show that the Regge poles and the associated greybody factor residues, for a massless scalar field theory, from which one can recover the black hole quasinormal complex frequencies as well as the fluctuations of the high energy absorption cross section, play naturally the role of critical parameters in the divergent behavior of the semiclassical deflection function. For very high frequencies, we finally recover the logarithmic part of the classical strong deflection limit, which clarifies analytically the fundamental link between quasinormal modes and strong gravitational lensing, suggested in recent works.
\end{abstract}


\maketitle

\section{Introduction}\label{intro}
Since the pioneering work of Darwin \cite{Darwin1959}, strong gravitational lensing by compact objects with a photon sphere, such as black holes (BH) and naked singularities, has been extensively studied, mainly motivated by indications of the presence of supermassive BH at the center of galaxies. It is well-established that strong lensing is related to photons passing close to the photon sphere and being able to circumvent multiple times the BH before eventually reaching an observer. Besides numerical studies \cite{VibhadraEllis1987,VibhadraKeeton2008}, the strong deflection limit (SDL) has been suggested as an analytical logarithmic approximation to describe the singular behavior of the strong lensing deflection function for photon impact parameters close to the critical impact parameter of the BH, allowing to describe relativistic images, magnifications and time delays. SDL has been rediscovered several times \cite{Luminet1979,Ohanian1987} and generalized to any asymptotically flat, static and spherically symmetric BH \cite{Bozza2002}, before being extended to Kerr BH \cite{Bozza2003,BozzaDeLucaScarpettaSereno2005,BozzaDeLucaScarpetta2006} or applied to alternative theories and string theory \cite{Eiroa2006,EiroaSendra2014,MukherjeeMajumdar2007}. For a review on gravitational lensing, we refer the reader to \cite{Perlick2004}.
On the other hand, it has been shown geometrically and semiclassically \cite{FerrariMashhoon1984,KokkotasSchmidt1999,CardosoMirandaBertiWitekZanchin2009,DecaniniFolacciJensen2003,DolanOttewill2009,DecaniniFolacci2010,DecaniniFolacciRaffaelli2010} that the weakly damped quasinormal modes (QNM) of a BH are also intimately related to its photon sphere and hence to the properties of this hypersurface, support of unstable circular null geodesics. It is then very important to point out the obvious fundamental role played by the photon sphere in both strong gravitational lensing and the physics of QNM, leading one to think, in a very natural way, of a link between these two phenomena, which are nothing else than different aspects of the more general scattering theory by BH \cite{FuttermanHandlerMatzner,FrolovNovikov}. This has been first underlined in \cite{StefanovYazadjievGyulchev2010}, then in \cite{WeiLiuGuo2011} and \cite{WeiLiu2014}, by formally comparing the strong lensing parameters, introduced in \cite{Bozza2002}, with the eikonal limit of the QNM complex frequencies and with the expression of the high energy absorption cross section. However an analytical proof is still missing. In this paper, using a semiclassical approach, we propose to unify the description of both phenomena and to understand analytically the link between the logarithmic divergence of the SDL and the QNM. More precisely, considering the simple case of a massless scalar field in a static and spherically symmetric background endowed with a photon sphere, and using a first-order WKB analysis \cite{SchutzWill1985,IyerWill1987} to solve the radial wave equation, we extract the real phase-shifts introduced by the Regge-Wheeler potential, from which we derive the singular part of the semiclassical deflection function \cite{FordWheeler1959,EisbergPorter1961}. Then, by analytically extending the results into the complex angular momentum (CAM) plane, we emphasize the explicit role of both the Regge poles and the associated greybody factor residues as critical parameters in the expression of our semiclassical deflection function. Finally, knowing that one can compute the QNM complex frequencies from the Regge poles and the related Regge trajectories in the CAM plane \cite{DecaniniFolacciRaffaelli2010}, this approach then allows us to prove on an analytical basis, the identification of the SDL parameters with the QNM complex frequencies in the eikonal limit, made in \cite{StefanovYazadjievGyulchev2010,WeiLiuGuo2011,WeiLiu2014}.

The paper is organized as follows. In Sec.~\ref{sec:2}, we recall some results about classical and semiclassical deflection functions in BH physics, and we introduce some notations and assumptions related to the SDL and to a scattering analysis within the CAM framework. In Sec.~\ref{sec:3}, we use the WKB approach developed in \cite{SchutzWill1985,IyerWill1987} that allows us to extract the real phase shifts introduced by the Regge-Wheeler potential, from which one can describe the singular logarithmic behavior of the semiclassical deflection function. In Sec.~\ref{sec:4}, we consider an analytical extension of the previous results into the complex angular momentum plane, first to show the natural transition between the physics of QNM and the SDL, and then to emphasize the role of the Regge poles and the associated greybody factor residues in the expression of the singular logarithmic part of the semiclassical deflection function. Considering the eikonal limit of our results, we recover the singular part of Bozza's deflection function. Finally, in Sec.~\ref{sec:5}, we extend our results to the example of the Schwarzschild-de Sitter BH, showing the effect of the cosmological constant in the singular part of the SDL, as well as the limits of our work. Throughout this paper, we shall use units such that $\hbar = c = G =1$ and we shall assume a harmonic time dependence $\exp(-i\omega t)$ for the massive scalar field.

\section{Generalities and notations}\label{sec:2}
We consider a static spherically symmetric four-dimensional spacetime with metric
\begin{equation} \label{metric_BH}
ds^2=-f(r)dt^2+\frac{dr^2}{f(r)}+r^2d\sigma^{2}.
\end{equation}
Here $d\sigma^{2}$ denotes the line element on the unit sphere $S^{2}$. In Eq.~(\ref{metric_BH}), we assume that $f(r)$ is a function of the usual radial Schwarzschild coordinate $r$ satisfying the following properties:

\begin{itemize}
\item There exists an interval $I=]r_h,+\infty[ \subset \bf{R}$ with $r_h>0$ such as $f(r)>0$ for $r \in I$.
\item $r_h$ is a simple root of $f(r)$, i.e. $f(r_h)=0$ and $f'(r_h)\neq 0$ and $f^{''}(r_h)>0$.
\item There exists a value $r_c \in I$ such as $r_c f'(rc)-2f(r_c)=0$ and $r_c^2 f''(r_c)-2f(r_c)<0$ 
\end{itemize}
These assumptions indicate that the spacetime considered has a single event horizon at $r_h$. Its exterior corresponds to $r \in I$ and it is endowed with a photon sphere (support of unstable circular null geodesics) located in $r_c$. If we want to work with an asymptotically flat spacetime, which will be needed to define an $S$-matrix or to consider Bozza's SDL, we need to impose the condition
\begin{equation}\label{Assump_f_2}
\underset{r \to +\infty}{\lim}f(r)=1.
\end{equation}
In the following, every quantity evaluated at $r=r_c$ will be noted with a subscript $c$, as in $f(r_c)=f_c$, and we introduce from now on the parameters $\eta_c$ defined by
\begin{equation}
\eta_c=\frac{1}{2}\sqrt{4f_c-2r_c^2f_c''}
\end{equation}
which represents, as suggested by the third of our previous assumptions, a measure of the instability of the circular orbits lying on the photon sphere. For more details about this parameter and its link with the Lyapunov exponent corresponding to the unstable circular orbits, we refer the reader to \cite{DecaniniFolacciRaffaelli2010} and references therein.

\subsection{The classical deflection function in the SDL}
In \cite{Bozza2002}, Bozza derived an expression of the deflection function in the SDL for static and spherically symmetric gravitational lenses with a photon sphere. The background is supposed to be asymptotically flat with the source and the object located ``far away'' from the BH, i.e. in the asymptotically flat region of the background. From the null geodesic equation of a free-falling massless particle of energy $E$ and angular momentum $L$, which are the two integrals of motion associated with the symmetries of our problem, one can easily obtain the integral expression of the deflection function
\begin{equation}\label{classical_deflection}
\Theta(r_0)=-\pi+2b\int_{r_0}^{+\infty}\frac{dr}{r^2}\left(1-f(r)\frac{b^2}{r^2}\right)^{-1/2},
\end{equation}
where $r_0$ is the classical turning point and $b=L/E=r_0/\sqrt{f(r_0)}$ is the impact parameter of the particle, which depends explicitly on $r_0$. We chose the same signs convention as in \cite{Bozza2002}. Then, the deflection function can be analytically approximated in two limiting cases: the weak deflection limit for photons passing at large distances from the BH, and the SDL. In the latter case, one can see from Eq.~(\ref{classical_deflection}) that, around the classical turning point $r_0$, the integral diverges if $r_0$ coincides with $r_c$, the maximum of the effective potential $V_\mathrm{eff}(r)=L^2f(r)/r^2$, i.e. the location of the photon sphere. This divergence is of logarithmic nature and a detailed treatment of this singular behavior can be found in \cite{Bozza2002}. One then obtains in the SDL case
\begin{equation}\label{deflection_Bozza}
\Theta(b)=-\pi+c_1\ln\left(\frac{b}{b_c}-1\right)+c_2+\underset{b \to b_c}{\mathcal{O}}\left(\frac{b}{b_c}-1\right),
\end{equation}
where $b_c=r_c/\sqrt{f_c}$ is the critical impact parameter of the BH, $c_1=-1/\eta_c$ is called the SDL coefficient in this paper, and $c_2$ a constant depending on $b_c$, $f_c$, $\eta_c$ and on the background geometry from $r_c$ to spatial infinity. It should be noted that the singular logarithmic behavior near a critical parameter, be it in a classical \cite{Darwin1959,Bozza2002} or a semiclassical framework \cite{FordWheeler1959,EisbergPorter1961}, is a universal feature for every deflection function of the form
\begin{equation}\label{general_deflection}
\Theta(b) \propto \int_{r_0}^{\infty} \frac{dr}{r^2}\left[1-W(r,b)\right]^{-1/2},
\end{equation}
where $b$ is a real parameter and $r_0$ is the classical turning point of the problem, such as $W(r_0,b)=1$, defining $b(r_0)$. If $r_0$ tends to $r_c$, with $r_c$ the maximum of the function $W(r,b)$, then a Taylor expansion of $W$ to the second order in $(r_0-r_c)$ allows to readily show that $\Theta$ indeed diverges logarithmically. 

\subsection{Partial waves and Regge-Wheeler equation}
In this section, we consider a massless scalar field $\Phi$ on a static and spherically symmetric background defined by Eq.~(\ref{metric_BH}). The Klein-Gordon wave equation, $\Box \Phi=0$, reduces, after separation of variables and introduction of the radial partial wave functions $\Phi_\ell(r)$ with $\ell=0,1,2, \dots$, to the Regge-Wheeler equation
\begin{equation}\label{RW}
\frac{d^2 \Phi_\ell}{dr_*^2} + \left[ \omega^2 - V_\ell(r)\right]\Phi_\ell=0.
\end{equation}
The variable $r_\ast=r_\ast(r)$ is the well-known tortoise coordinate defined, for $r \in I$, by the relation $dr_\ast/dr=1/f(r)$ which provides a bijection from $I$ to $]-\infty,+\infty[$. In Eq.~(\ref{RW}), $V_\ell(r)$ is the Regge-Wheeler potential associated with the massless scalar field
\begin{equation}\label{RWPotscalar}
V_\ell(r)=f(r) \left[ \frac{\ell(\ell+1)}{r^2}+\frac{1}{r}f'(r)\right].
\end{equation}
It should be noted that $\lim_{r \to r_h} V_\ell(r)=0$ and $\lim_{r \to +\infty} V_\ell(r)=0$. Therefore the solutions of the radial equation (\ref{RW}) have an $\exp(\pm i \omega r_{\ast})$ behavior at the horizon and at infinity. In other words, for a given angular momentum index $\ell$, a general solution of the Regge-Wheeler equation (\ref{RW}), satisfies the following asymptotic behaviors:
\begin{equation}\label{bc1}
\Phi_{\omega,\ell}^{(-\infty)} (r_\ast) \underset{r_\ast \to -\infty}{\sim} e^{-i\omega r_\ast }
\end{equation}
which is a purely ingoing wave at the event horizon, and which has the following general expression at spatial infinity $r_\ast \to +\infty$
\begin{equation}\label{bc2}
\Phi_{\omega,\ell}^{(+\infty)}(r_\ast)\underset{r_\ast \to +\infty}{\sim} A_{in}e^{-i\omega r_\ast}+A_{out}e^{+i\omega r_\ast}.
\end{equation}
Moreover, considering the wronskian of two linearly independent solutions of Eq.~(\ref{RW}) at $r_\ast=\pm\infty$, like for example Eq.~(\ref{bc2}) and its complex conjugate, one obtains
\begin{equation}\label{Wrsk}
1+\left|A_{out}\right|^2=\left|A_{in}\right|^2.
\end{equation}
Finally, introducing the more usual transmission and reflection amplitudes $T_\ell$ and $R_\ell$ defined by
\begin{equation}\label{TransmissionReflection}
T_\ell=\frac{1}{A_{in}} \quad \mathrm{and} \quad R_\ell=\frac{A_{out}}{A_{in}},
\end{equation}
we can rewrite Eq.~(\ref{Wrsk}) in the more familiar form
\begin{equation}
\left|T_\ell\right|^2+\left|R_\ell\right|^2=1.
\end{equation}
It is worth recalling that the transmission amplitudes $T_\ell(\omega)$ permit us to construct the greybody factors (the absorption probabilities by the BH for scalar particles with energy $\omega$ and angular momentum $\ell$). They are defined according to
\begin{equation}\label{Greybodyfactors}
\Gamma_\ell(\omega)= |T_\ell(\omega)|^2=\frac{1}{|A_{in}|^2},
\end{equation}
while the reflection coefficients $R_\ell(\omega)$ are linked to the $S$-matrix elements, noted $S_\ell(\omega)$, of our spherically symmetric problem through
\begin{equation}\label{Smatrix}  
S_\ell(\omega)=(-1)^{\ell+1}R_\ell(\omega)=(-1)^{\ell+1}\frac{A_\mathrm{out}}{A_\mathrm{in}}.
\end{equation}
From Eqs.~(\ref{bc1}) and (\ref{bc2}), a QNM, which is defined as a purely ingoing wave at the event horizon and a purely outgoing wave at infinity, corresponds to $A_{in}=0$. It has been shown in \cite{DecaniniFolacciJensen2003,DecaniniFolacci2010,DecaniniFolacciRaffaelli2010} (and references therein) that the $S$-matrix permits to analyze the resonant aspects of the considered BH as well as to construct the form factor describing the scattering of a monochromatic scalar wave.

\subsection{The semiclassical deflection function}
The deflection function is an important notion in scattering problems. Ford and Wheeler gave a very detailed description in their seminal work on semiclassical scattering \cite{FordWheeler1959}. Over the past few decades, it has been shown that their formalism can easily be extended to BH physics. In particular, the semiclassical deflection function can be defined from the $S$-matrix elements $S_\ell$, with $\ell$ supposed to take on real continuous values. It corresponds to the angle $\Theta(\ell)$ by which a partial wave, with angular momentum $\ell$, is scattered by a BH. It reads
\begin{equation}\label{semiclassical_deflection_def1}
\Theta(\ell)\myeq\frac{d}{d\ell}\mathrm{arg}\left[S_\ell\right].
\end{equation}
With Eq.~(\ref{Smatrix}) and taking into account the same signs convention for the deflection function as in \cite{Bozza2002}, Eq.~(\ref{semiclassical_deflection_def1}) can also be written as
\begin{equation}\label{semiclassicaldeflection_def}
\Theta(\ell)=-\pi+\frac{d}{d\ell}\mathrm{arg}\left[\frac{A_\mathrm{out}}{A_\mathrm{in}}\right].
\end{equation}
It should be noted that it is often more convenient to write the $S$-matrix elements as $S_\ell=e^{2i\delta_\ell}$, where $\delta_\ell \in \mathbb{C}$ are the related phase shifts, when one has to work with the scattering amplitude. In such case, the deflection function reads
\begin{equation}\label{semiclassical_deflection_def2}
\Theta(\ell)=2\frac{d}{d\ell}\mathrm{Re}\left[\delta_\ell\right].
\end{equation}
To solve Eq.~(\ref{semiclassical_deflection_def2}), Ford and Wheeler suggested to consider phase shifts $\delta_\ell$ approximated by a one turning point WKB definition, $\delta^\mathrm{WKB}_\ell$ \cite{FordWheeler1959,EisbergPorter1961}. 
One then recovers, from the right hand side of Eq.~(\ref{semiclassical_deflection_def2}), the deflection function of classical physics, i.e. Eq.~(\ref{general_deflection}), by introducing an effective impact parameter for each partial waves
\begin{equation}\label{semiclassical_impactparameter}
b=\frac{\ell+1/2}{\omega}.
\end{equation}
However, the Regge-Wheeler potential has two turning points and such a procedure can't be used in our context. It should be noted that in \cite{GlampedakisAndersson2001}, the authors suggested to reduce the BH scattering problem to a one turning point problem by considering the limit $\ell/\omega \rightarrow +\infty$. Indeed, in this limit, the scattering is mainly due to the outer turning point, allowing to derive a one turning point WKB approximation for the phase shifts (similiar considerations, for the acoustic BH, can be found in \cite{DolanOliveiraCrispino2009} where a Born approximation is equivalently used). The expression obtained for the deflection function eventually matches the classical expression given by Eq.~(\ref{classical_deflection}), albeit only at large distances, as expected from the $\ell/\omega$ assumption. However, some years ago, Schutz, Will and Iyer \cite{SchutzWill1985,IyerWill1987} developed a WKB analysis which is particularly well-adapted to BH physics, allowing us to treat properly the two turning points problem near the maximum of the Regge-Wheeler potential. Actually, this WKB analysis has been the starting point used in the semiclassical study of the weakly damped QNM in \cite{DecaniniFolacci2010,DecaniniFolacciRaffaelli2010} within the CAM theory framework, and we will also use it in the following section.

\section{WKB analysis: Phase shifts and Regge poles}\label{sec:3}
\subsection{First-order WKB analysis: induced phase shifts}\label{1stWKBintro}

The classical deflection function approximated in the SDL by Eq.~(\ref{deflection_Bozza}) has been obtained in the geometrical, i.e. eikonal, framework of General Relativity. Therefore, to recover the logarithmic singular behavior from semiclassical methods, it should be reasonable to consider a first-order WKB approximation. To do so, we follow carefully the WKB analysis introduced in \cite{SchutzWill1985,IyerWill1987} mentioned in the previous section, close to the maximum of the Regge-Wheeler potential. We start by writing the general solutions (see Eq.~(3.5) of \cite{IyerWill1987}) of the scalar field at the horizon and at infinity in the following form
\begin{eqnarray}
\Phi_{\ell,\omega}^{(-\infty)}(r_\ast) &\underset{r_\ast \to -\infty}{\sim}& Z^\mathrm{III}_\mathrm{in} \Phi^\mathrm{III}_{+} + Z^\mathrm{III}_\mathrm{out} \Phi^\mathrm{III}_{-}\nonumber\\
\Phi_{\ell,\omega}^{(+\infty)}(r_\ast) &\underset{r_\ast \to +\infty}{\sim}& Z^\mathrm{I}_\mathrm{in} \Phi^\mathrm{I}_{-} + Z^\mathrm{I}_\mathrm{out} \Phi^\mathrm{I}_{+},
\end{eqnarray}
where the superscript ``$\mathrm{I}$'' denotes the region from the horizon to the inner turning point, ``$\mathrm{III}$'' is the region from the outer turning point to infinity and the subscripts ``$\mathrm{in}$'' and ``$\mathrm{out}$'' represent respectively the ingoing and the outgoing components of the wavefunction $\Phi$. For more details about the notations, we refer the reader to \cite{IyerWill1987}. In this paper, we introduce a function $\mathcal{E} \in \mathbb{R}$ (with $i\mathcal{E}(\ell,\omega)=(\nu+1/2)$ in Iyer and Will's notations), which reads at first WKB order:
\begin{equation}\label{EWKB}
\mathcal{E}(\ell,\omega)=\frac{\omega^2-V_{0}(\ell)}{\sqrt{-2V_{0}^{''}(\ell)}},
\end{equation}
where $V_0(\ell)$ and $V_0^{''}(\ell)$ are respectively the Regge-Wheeler potential and its second derivative, evaluated at the maximum $(r_\ast)_0$ of $V_\ell(r_\ast)$. We finally obtain a connection formula (see Eq.~(3.33) of \cite{IyerWill1987}), written as:
\begin{widetext}
\begin{equation}\label{connection_formula}
\begin{pmatrix}
Z^\mathrm{III}_\mathrm{out}  \\
  Z^\mathrm{III}_\mathrm{in}\\
\end{pmatrix}
=
\begin{pmatrix}
-ie^{-\pi\mathcal{E}(\ell,\omega)} & e^{-i\phi(\ell,\omega)}(1+e^{-2\pi\mathcal{E}(\ell,\omega)})^{1/2}  \\
  e^{i\phi(\ell,\omega)}(1+e^{-2\pi\mathcal{E}(\ell,\omega)})^{1/2} & ie^{-\pi\mathcal{E}(\ell,\omega)}  \\
\end{pmatrix}
\begin{pmatrix}
Z^\mathrm{I}_\mathrm{out}  \\
  Z^\mathrm{I}_\mathrm{in}\\
\end{pmatrix},
\end{equation}
\end{widetext}
where $\phi(\ell,\omega) \in \mathbb{R}$ is defined by
\begin{eqnarray}\label{WKBphase}
\phi(\ell,\omega)&=&\mathcal{E}(\ell,\omega) - \mathcal{E}(\ell,\omega) \ln\left|\mathcal{E}(\ell,\omega)\right|\nonumber\\ 
&& \qquad \qquad +\arg\left[\mathrm{\Gamma}\left(i\mathcal{E}(\ell,\omega)+1/2\right)\right],
\end{eqnarray}
and $\mathrm{\Gamma}$ is the gamma function, not to be confused with $\Gamma_\ell$, the greybody factors introduced previously.
With Eqs.~(\ref{bc1}) and (\ref{bc2}), we have $Z^\mathrm{I}_\mathrm{in}/Z^\mathrm{III}_\mathrm{out}=A_\mathrm{in}$, $Z^\mathrm{I}_\mathrm{out}/Z^\mathrm{III}_\mathrm{out}=A_\mathrm{out}$ and of course, for a BH $Z^\mathrm{III}_\mathrm{in}=0$.  
In particular, from Eq.~(\ref{connection_formula}) and because the functions $\mathcal{E}$ and $\phi$ are real, one has
\begin{equation}\label{phase_deflection}
\arg\left[\frac{A_\mathrm{out}}{A_\mathrm{in}}\right]=-\frac{\pi}{2}-\phi(\ell,\omega).
\end{equation}
As a consequence, according to Eqs.~(\ref{semiclassicaldeflection_def}), (\ref{WKBphase}) and (\ref{phase_deflection}), we conclude that the main contribution describing the logarithmic divergent behavior of the deflection function $\Theta^\mathrm{sing}$ is
\begin{equation}\label{ThetaSing}
\Theta^{\mathrm{sing}}(\ell,\omega) \sim \frac{d\mathcal{E}(\ell,\omega)}{d\ell}\ln\left|\mathcal{E}(\ell,\omega)\right|.
\end{equation}
Here, the function $\mathcal{E}(\ell,\omega)$ plays obviously an important role in the semiclassical computation of $\Theta^{\mathrm{sing}}$. Actually, $\mathcal{E}(\ell,\omega)$ appears to be the key function for the purpose of this paper because, as we will see next, it is also behind the computation of the Regge poles, from which one is able to describe semiclassically the QNM complex frequencies.

\subsection{Regge poles and Quasinormal modes} 
In this paper, we consider the description of QNM through the CAM theory \cite{DecaniniFolacciRaffaelli2010} which is the most relevant approach for our purpose. Within the CAM framework, the QNM are considered as generated by a family of ``surface waves'' lying close to the BH photon sphere at $r_c$, and the corresponding QNM complex frequencies are interpreted as Breit-Wigner-type resonances. To work in the CAM plane, we transform the angular momentum $\ell$ into a complex variable $\lambda=\ell+1/2$, for every $\omega>0$. The functions $T_\ell$, $\Gamma_\ell$ defined by (\ref{Greybodyfactors}), $R_\ell$ and $S_\ell$, introduced in Eq.~(\ref{Smatrix}), as well as $\mathcal{E}(\ell,\omega)$ can be analytically extended into the complex $\lambda-$plane, giving respectively the analytical extensions $T_{\lambda-1/2}$, $\Gamma_{\lambda-1/2}$, $R_{\lambda-1/2}$, $S_{\lambda-1/2}$ and $\mathcal{E}(\lambda-1/2,\omega)$. The Regge poles, from which one can recover the QNM complex frequencies, are the poles in the CAM plane of the function $T$ (an equivalently of $\Gamma$, $R$ and $S$), i.e. satisfying the condition $A_\mathrm{in}=0$. In particular, we have from Eqs.~(\ref{Greybodyfactors}) and (\ref{connection_formula})
\begin{equation}\label{Greybodyfactors_Analytic} 
\Gamma_{\lambda-1/2}(\omega)=\frac{1}{1+\exp[-2\pi\mathcal{E}(\lambda-1/2,\omega)]}.
\end{equation}
The Regge poles are then the solutions $\lambda_n(\omega)$ of the equation
\begin{equation}\label{RP_phase}
\mathcal{E}(\lambda_n-1/2,\omega)= i(n-1/2) \quad \mathrm{with}
\quad n \in \mathbb{N}\setminus\lbrace{0\rbrace}
\end{equation}
(here we consider the poles lying in the first quadrant of the complex $\lambda$-plane, see \cite{DecaniniFolacciRaffaelli2010}). One can also easily prove that the corresponding residues are given by
\begin{equation}\label{residues}
\gamma_n(\omega)=\frac{1/2\pi}{[d\mathcal{E}(\lambda-1/2,\omega)/d\lambda]|_{\lambda=\lambda_n(\omega)}}.
\end{equation}
However, in the following, in order to simplify our notations, we introduce $\rho_n(\omega)$ such as
\begin{equation}\label{rho_def}
\rho_n(\omega) \myeq \frac{1}{2\pi\gamma_n(\omega)}=\left.\frac{d\mathcal{E}(\lambda-1/2,\omega)}{d\lambda}\right|_{\lambda=\lambda_n(\omega)}.
\end{equation}                                         
Using Eq.~(\ref{EWKB}) and other higher WKB orders \cite{DolanOttewill2009,DecaniniFolacciRaffaelli2010,DecaniniFolacciRaffaelli2011}, in the range of $\lambda$ such as $|\lambda| \gg 1$, $\mathrm{Re}\,\lambda \gg \mathrm{Im}\,\lambda$ and $|d/d\omega\,\mathrm{Re}\,\lambda_n(\omega)| \gg |d/d\omega\,\mathrm{Im}\,\lambda_n(\omega)|$, one can finally write their respective high frequency expressions
\begin{widetext}
\begin{subequations}\label{RPResiduesRho}
\begin{equation}\label{ReggePoles}
\lambda_n(\omega)=\frac{r_c}{\sqrt{f_c}}\omega+i\eta_c(n-1/2)+\frac{a_n/2}{(r_c/\sqrt{f_c})}\frac{1}{\omega}+\underset{\omega  \to
+\infty}{\mathcal{O}}\left(\frac{1}{[(r_c/\sqrt{f_c})\omega]^2}\right)
\end{equation}
\begin{equation}\label{Residues}
\gamma_n(\omega)=-\frac{\eta_c}{2\pi}+i\frac{a_c(n-1/2)/2\pi}{(r_c/\sqrt{f_c})}\frac{1}{\omega}+\underset{\omega  \to
+\infty}{\mathcal{O}}\left(\frac{1}{[(r_c/\sqrt{f_c})\omega]^2}\right)
\end{equation}
\begin{equation}\label{rho_DA}
\rho_n(\omega)=-\frac{1}{\eta_c}-i\frac{a_c(n-1/2)}{\eta_c^2(r_c/\sqrt{f_c})}\frac{1}{\omega}+\underset{\omega  \to
+\infty}{\mathcal{O}}\left(\frac{1}{[(r_c/\sqrt{f_c})\omega]^2}\right),
\end{equation}
\end{subequations}
\end{widetext}
where $a_n$ and $a_c$ are defined in the aforementioned references. Even though in this paper we restrict ourselves to the first WKB order, it is still interesting to point out through Eqs.~(\ref{RPResiduesRho}), the higher order terms that should contribute to strong lensing, for a future study.

\noindent From the analytical extension into the complex $\lambda-$plane and the Regge poles expressions, one can have a lot of information about the weakly damped QNM. Indeed, in \cite{DecaniniFolacci2010} the authors showed that, for a massless scalar field propagating on a static and spherically symmetric background, the complex frequencies of the weakly damped QNM can be obtain analytically from the Regge trajectories, i.e., from the curves traced out in the CAM plane by the Regge poles as a function of the frequency $\omega$, because of the dual structure of the S-matrix elements, function of both the frequency $\omega$ and the angular momentum index $\ell$. In particular, in the complex $\omega$-plane, we denote the QNM complex frequencies as $\omega_{\ell n}=\omega^{(o)}_{\ell n}- i\Gamma_{\ell n}/2$ with $\ell \in \mathbb{N}$, $n\in \mathbb{N}\setminus\lbrace{0\rbrace}$, $\omega^{(o)}_{\ell n}>0$ and $\Gamma_{\ell n}>0$. The Regge trajectories then allow us to interpret the Regge poles as surface waves propagating close to the photon sphere, with $\mathrm{Re}\,\lambda_n(\omega)$ providing the dispersion relation of the $n^{th}$ surface wave and $\mathrm{Im}\,\lambda_n(\omega)$ corresponding to its damping. Finally, the excitation frequencies $\omega^{(o)}_{\ell n}$ and the widths $\Gamma_{\ell n}/2$ of the resonances are given by the semiclassical relations
\begin{subequations} \label{sc12}
\begin{equation}\label{sc1}
\mathrm{Re}  \, \lambda_n \left(\omega^{(0)}_{\ell n} \right)= \ell
+ 1/2   \qquad \ell \in \mathbb{N},
\end{equation}
and
\begin{equation}\label{sc2} \frac{\Gamma _{\ell n}}{2}= \left.  \frac{\mathrm{Im} \, \lambda_n
(\omega )}{d/d\omega \, \ \mathrm{Re} \, \lambda_n (\omega )  }
\right|_{\omega =\omega^{(0)}_{\ell n}}.
\end{equation}
\end{subequations}
It should also be noted that with the Regge poles and the greybody factor residues, one is also able, from an analytical extension into the complex $\lambda$-plane of the greybody factors, and using CAM techniques on the partial wave series expression of the absorption cross section, to describe analytically the fluctuations of the high energy absorption cross section for a massless scalar field propagating in a static and spherically symmetric
BH with a photon sphere. Although this point is important to understand \cite{WeiLiuGuo2011}, we will not develop this aspect here and refer the reader to \cite{DecaniniEspositoFareseFolacci2011} for more details, keeping in mind that the Regge poles and the greybody factor residues are the fundamental quantities common to the physics of QNM, to the fluctuations of the high energy absorption cross section and to the semiclassical deflection function, as we will see in the following.

\noindent Finally, to conclude this section, let us recall the key role of the function $\mathcal{E}(\ell,\omega)$ in the description of both the (singular part of the) semiclassical deflection function and the QNM through the study of the Regge poles. In the next section, we perform the natural transition from the physics of QNM to the SDL and, at the eikonal limit, we recover the logarithmic part in Eq.~(\ref{deflection_Bozza}) of Bozza's deflection function. 

\section{From QNM to SDL}\label{sec:4}
In Sec.~\ref{1stWKBintro}, the function $\mathcal{E}(\ell,\omega)$ and therefore the phase $\phi(\ell,\omega)$ must be real functions so that one can define a deflection function $\Theta(\ell,\omega)$ with Eqs.~(\ref{phase_deflection}) and (\ref{ThetaSing}). Therefore, in the analytical extension into the complex $\lambda$-plane, we claim that only the real part of the function $\mathcal{E}(\lambda-1/2,\omega)$ contributes to the singular logarithmic behavior of Eq.~(\ref{ThetaSing}). More precisely, it is suggested that the SDL should be described by complex angular momentum values close to the Regge poles, leading us to perform a first-order series expansion of $\mathcal{E}(\lambda-1/2)$ around a pole, the zeroth order corresponding to the physics of QNM. In the range of $\lambda$ used above, we will then consider as a first approximation that
\begin{equation}
\mathcal{E}(\ell,\omega) \sim \mathrm{Re}\left[\mathcal{E}(\lambda-1/2,\omega)\right]. 
\end{equation}
For $\lambda$ close to a Regge pole $\lambda_n(\omega)$, and taking into account Eqs.~(\ref{RP_phase}) and (\ref{rho_DA}), we obtain at first-order in $\lambda$
\begin{equation}\label{TaylorRP_phase}
\mathcal{E}(\lambda-1/2,\omega) \sim i(n-1/2)+[\lambda-\lambda_n(\omega)]\rho_n(\omega),
\end{equation}
which gives for $\mathcal{E}(\ell,\omega)$
\begin{equation}\label{closephotonsphere}
\mathcal{E}(\ell,\omega)\sim\mathrm{Re}\left\{[\lambda-\lambda_n(\omega)]\rho_n(\omega)\right\}
\end{equation}
and, as a consequence
\begin{eqnarray}
\frac{d\mathcal{E}(\ell,\omega)}{d\ell}&\sim&\frac{d}{d\mathrm{Re}(\lambda-1/2)}\mathrm{Re}\left\{[\lambda-\lambda_n(\omega)]\rho_n(\omega)\right\}\nonumber\\
&=&\mathrm{Re}\left[\rho_n(\omega)\right].
\end{eqnarray}
Therefore, the singular part of the deflection function, Eq.~(\ref{ThetaSing}), for $\lambda$ close to $\lambda_n(\omega)$, reads in its most compact form
\begin{equation}\label{deflection_compact}
\Theta^{\mathrm{sing}}_n(\ell,\omega)=\mathrm{Re}\left[\rho_n(\omega)\right]\ln\left|\mathrm{Re}\left\{\left[\lambda-\lambda_n(\omega)\right]\rho_n(\omega)\right\}\right|.
\end{equation}
Hence, for $\lambda=\lambda_n(\omega)$, $\Theta^\mathrm{sing}$ diverges, which corresponds to the capture of the photon on the photon sphere from the strong lensing point of view, while it describes the generation of surface waves from the QNM point of view.

\noindent Before going any further, it should be noted that the deflection function depends on the angular momentum $\ell=\mathrm{Re}\lambda-1/2$ as in usual semiclassical physics \cite{FordWheeler1959,EisbergPorter1961}, but also on the Regge pole $\lambda_n(\omega)$ and therefore on both its label $n$ and the energy $\omega$ of the massless scalar field. If the deflection function in Eq.~(\ref{deflection_compact}) describes strong gravitational lensing in the SDL, then it would be natural to expect, as opposed to classical General Relativity results, at least from its singular part, new fine effects such as ``iridescence'' due to the frequency dependence, as well as an additional substructure due to the labeling of the poles. However, as noted in the previous section, we will focus in the following, on the first WKB order of the Regge poles and the associated residues, in order to recover the singular logarithmic part of the deflection function derived by Bozza. Then, some of these new features will not be explicit in this paper but will be studied in the future, by considering higher orders in the WKB analysis. Therefore, in the eikonal limit of Eqs.~(\ref{RPResiduesRho}) in Eq.~(\ref{deflection_compact}), we finally recover Bozza's logarithmic singular part
\begin{equation}\label{ThetaSingBozza}
\Theta^{\mathrm{sing}}(b,\omega)=-\frac{1}{\eta_c}\ln\left(\frac{b}{b_c}-1\right)-\frac{1}{\eta_c}\ln\left(\frac{b_c\omega}{\eta_c}\right).
\end{equation}
It should be noted that we do not recover the $c_2$ term in Eq.~(\ref{deflection_Bozza}). This is mainly due to the asymptotic ``plane'' wave behavior (in the tortoise coordinate $r_\ast$) in Eqs.~(\ref{bc1}) and (\ref{bc2}), making the problem equivalent to a ``flat'' spacetime semiclassical problem \cite{EisbergPorter1961}, unlike Bozza's $c_2$ coefficient which depends on the background geometry, from $r_c$ to spatial infinity. Actually, the $c_2$ term seems to not be linked to any QNM-related quantity in \cite{StefanovYazadjievGyulchev2010} either. In other words, the WKB analysis performed in the previous section, allows to describe, as expected, ``local'' physics, i.e. near the maximum of the Regge-Wheeler potential, or ``intrinsic'' features of the semiclassical SDL, i.e. independently of the respective distances between the source, the lens and the observer, as opposed to the ``global'' $c_2$ term. Moreover, as in usual semiclassical study \cite{FordWheeler1959,EisbergPorter1961}, we inherited an unavoidable, intrinsic $\omega$-dependent logarithmic term. Even though, for very high frequency, the variation of $\Theta^\mathrm{sing}$ with respect to $\omega$, which goes like $\omega^{-1}$ (at fixed impact parameter $b$), is of course negligible, it could affect the observations for a lower range of frequency. Therefore, for lower frequencies and at fixed impact parameter $b$, one could a priori observe (with future technologies) iridescence due to this $\omega$-dependent term. But what do we mean by lower frequencies? Is there any lower bound from which iridescence could be measured through $\Theta$? Actually, if Eq.~(\ref{ThetaSingBozza}) is seen as a semiclassical correction of the singular logarithmic part of Eq.~(\ref{deflection_Bozza}), then one shall compare the intrinsic $\omega$-dependent term with the ``global'' $c_2$ term, if the latter is also a measurable quantity. More precisely, according to \cite{Bozza2002}, for a Schwarzschild BH of mass $M=10 M_\odot$, the $\omega$-term is of the order of $c_2$ for $\omega \gtrsim 6.10^4 \mathrm{Hz}$ and for a supermassive Schwarzschild BH of mass $M=10^6 M_\odot$, which will be from an experimental point of view the objects of interest in the near future, this occurs for $\omega \gtrsim 10^{-1} \mathrm{Hz}$, which are, in both cases only 10 times greater than the QNM frequency scales.

\section{Effect of the cosmological constant on strong gravitational lensing: the Schwarzschild-de Sitter BH case}\label{sec:5}
In this section, we would like to apply our results to a 4-dimensional Schwarzschild-de Sitter (SdS) BH of mass $M$, of cosmological interest, which describes a non asymptotically flat spacetime that is therefore outside the scope of both Bozza's classical approach and, a priori, also out of the scope of the S-matrix formalism. In the SdS BH case, the function $f(r)$ reads
\begin{equation}\label{SdSmetric}
f(r)=1-\frac{2M}{r}-\frac{r^2}{L^2},
\end{equation}
where $L$ is a length linked to the cosmological constant $\Lambda$ through $\Lambda=3/L^2$. If
\begin{equation}
0<\frac{27M^2}{L^2}<1,
\end{equation}
then the equation $f(r)=0$ has only two real roots $r_h$ and $r_{Co}$, such as $0<r_h<r_{Co}$. They correspond respectively to the locations of the BH horizon and of the cosmological horizon. For any $r \in [r_h,r_{Co}]$, we have $f(r)>0$. Because the SdS spacetime is not asymptotically flat and $f(r)$ is positive only in $[r_h,r_{Co}]$, Bozza's classical approach to SDL and the S-matrix formalism of CAM theory, can't be used. However, it is still possible to circumvent this difficulty:
\begin{itemize}
\item First, because of Eq.~(\ref{SdSmetric}), the Regge-Wheeler potential $V_\ell(r)$, in Eq.~(\ref{RWPotscalar}), satifies now the boundary condition limits: $\lim_{r \to r_h} V_\ell(r)=0$ and $\lim_{r \to r_{Co}} V_\ell(r)=0$. Since the tortoise coordinate $r_\ast(r)$ still provides a bijection from $[r_h,r_{Co}]$ to $[-\infty,+\infty]$, it is then formally possible to define a ``kind of'' S-matrix from Eqs.~(\ref{bc1}) and (\ref{bc2}) and then to consider the associated deflection function defined by Eq.~(\ref{semiclassicaldeflection_def}).

\item Second, the structure of this S-matrix in the complex $\omega$-plane and in the $\lambda$-plane allows us to consider respectively the spectrum of the quasinormal complex frequencies and the spectrum of the Regge poles.

\item It becomes then possible to link semiclassically the two spectra with Eqs.~(\ref{sc12}) by extending a formalism introduced by Sommerfeld \cite{Sommerfeld} instead of the usual approach to scattering \cite{Watson1918}. In few words, this can be done by constructing the diffractive part of the Feynman propagator for the scalar field from the Regge modes as, for example, in \cite{DecaniniFolacci2009}.
\end{itemize}
For the 4-dimensional SDS BH, the location of the photon sphere $r_c$ and the instability parameter $\eta_c$ are given by
\begin{eqnarray}
&&r_c=3M \in [r_h,r_{Co}]\nonumber\\
&&\eta_c=1.
\end{eqnarray}
Moreover, it has been shown \cite{DecaniniFolacciRaffaelli2010} that, in the eikonal limit, the Regge poles are given by
\begin{equation}\label{RPSdS}
\lambda_n(\omega)=\frac{3\sqrt{3}M}{\sqrt{1-27M^2/L^2}}\omega + \underset{\omega \to +\infty}{\mathcal{O}}\left(\omega^{-1}\right).
\end{equation}
It is very interesting to note that while $r_h$ and $\eta_c$ are independent of the cosmological constant, the Regge poles $\lambda_n(\omega)$ are clearly dependent on $\Lambda$ through $L$. As a consequence, if the source and the observer are not in larger distances from the lens than $r_{Co}$ and using Eqs.~(\ref{deflection_compact}) and (\ref{RPSdS}), then the cosmological constant should also affect, from a semiclassical framework, the strong gravitational lensing in agreement with the analysis of \cite{RindlerIshak2007}. More precisely, due to the large value of $L$, on can perform a Taylor expansion of Eq.~(\ref{deflection_compact}) which then reads as the singular part of the deflection function for the usual 4-dimensional Schwarzschild BH case, plus a perturbative, intrinsic $L$-dependent term
\begin{equation}
\Theta^{\mathrm{sing}}(b,L,\omega)=\Theta^{\mathrm{sing}}_{\mathrm{Schw}}(b,\omega)+\frac{27M^2}{2L^2\left(\frac{b}{3\sqrt{3}M}-1\right)},
\end{equation}
where $\Theta^{\mathrm{sing}}_{\mathrm{Schw}}(b,\omega)$ is
\begin{equation}
\Theta^{\mathrm{sing}}_{\mathrm{Schw}}(b,\omega)=-\ln\left(\frac{b}{3\sqrt{3}M}-1\right)-\ln\left(3\sqrt{3}M\omega\right).
\end{equation}
It should be noted that the intrinsic $L$-dependent term and the intrinsic Schwarzschild term have opposite signs, as expected. However, the contribution of $\Lambda$ to the singular part of $\Theta$ is very small. Indeed, we know from cosmology that $\Lambda \approx 10^{-52} m^{-2}$. Therefore, for BH of mass $M=10M_\odot$ to $10^6M_\odot$, the ratio $M^2/L^2$ is of the order of $10^{-44}$ to $10^{-34}$, so ridiculously small that one would be led to consider huge BH of masses $M \approx 10^{20}M_\odot$ to observe the first effects of this local contribution. This is precisely where our semiclassical analysis reaches its limit. As a local study (near the maximum of the Regge-Wheeler potential), it does not bring any information about the global contributions (distant regions from the BH) in strong lensing, i.e. about terms that should also depend for example, on the distances between the source, the lens and the observer. More precisely, for the SdS case, as suggested in \cite{RindlerIshak2007,IshakRindlerDossett2010,Schucker2009}, those terms would be at the origin of more reasonable effects of the cosmological constant in gravitational lensing, and hence shall be the purpose of a future work on the semiclassical SDL.

\section{Conclusion}
The approach developed in this paper is an attempt to describe two important features of the scattering theory by BH originating close to the photon sphere, namely the strong gravitational lensing and the QNM, in a semiclassical unified scheme. More precisely, we considered a massless scalar field theory on a 4-dimensional, static and spherically symmetric geometry endowed with a photon sphere. We used a first-order WKB approximation adapted to the two turning points problem of BH physics, near the maximum of the Regge-Wheeler potential, i.e. close to the photon sphere, to solve the associated radial wave equation. From the WKB real phase shifts, we were then able to evaluate the divergent part of the semiclassical deflection function, emphasizing the key role of the function $\mathcal{E}(\ell,\omega)$ which, from an analytical extension into the complex $\lambda$-plane of CAM theory, allows also to compute the Regge poles and hence, to obtain the QNM complex frequencies. Therefore, in this semiclassical framework, we were finally able to prove what was still a formal identification between the main quantities appearing in Bozza's SDL and in the eikonal limit of QNM frequencies. At first-order, our analysis also sheds a new light on new effects in strong lensing, like iridescence and the role of the cosmological constant. However, our main result is an intrinsic feature of the semiclassical SDL, derived from a ``local'' analysis (restricted to the near photon sphere region), independent of the distances between the source, the lens and the observer. Therefore, it does not bring any information about what is happening in more extended regions of spacetime, where some interesting effects seems to occur as well, especially due to the cosmological constant. Finally, it is important to note that our analysis is only a first step to understand strong gravitational lensing by BH, from a semiclassical perspective. In the future, it would be interesting first to extend the WKB analysis to higher orders, to see the effects of the non-linear frequency dependence of the Regge poles and the effects of their labeling, as far as it could be relevant to observations and then, to understand, still in a semiclassical framework, the origin of the source/lens/observer distances-dependent terms, which seems also to play an important role.




\end{document}